\documentclass[11pt,twocolumn, notitlepage]{article}
\usepackage[utf8]{inputenc}
\usepackage{authblk}
\usepackage{cite}
\usepackage{url}
\usepackage{xcolor}
\usepackage{graphicx}
\usepackage{amsmath}
\usepackage{amssymb}
\usepackage{wasysym}
\usepackage{bm}
\usepackage{adjustbox}
\usepackage[T1]{fontenc}
\usepackage{babel}
\usepackage{siunitx}
\usepackage{pgfplots}
\usepackage{float}
\usepackage{caption}
\usepackage{subcaption}
\usepgfplotslibrary{groupplots}
\pgfplotsset{compat=newest}
\pgfplotsset{
  every axis plot/.append style={very thick},
}
\pgfplotsset{width=10cm}
\usepackage[margin=1in]{geometry}
\graphicspath{./images/}

\title{Effect of molecular rotation and concentration on the adsorption of pentacene molecules on two-dimensional monolayer transition metal dichalcogenides}
\author[1,*]{E. Black}
\author[1]{J. M. Morbec}
\affil[1]{School of Chemical and Physical Sciences, Keele University, Keele ST5 5BG, U.K.}
\affil[*]{Corresponding author: e.j.black@keele.ac.uk}
\date{}
\begin{document}
\twocolumn[
\begin{@twocolumnfalse}
\maketitle
\begin{abstract}
Heterostructures composed of pentacene (PEN) molecules and transition metal dichalchogenides (TMDs) are promising materials for small, flexible and lightweight photovoltaic devices and various other optoelectronic applications. The effects of changing concentration and orientation of adsorbed pentacene molecules on two-dimensional monolayer substrates of TMDs, namely  MoS$_2$, MoSe$_2$, WS$_2$ and WSe$_2$, were investigated using first-principles calculations based on density functional theory. We examined the structural and electronic properties of the corresponding PEN/TMD heterostructures and compared these between differing pentacene concentrations and the orientations of pentacene with respect to the underlying substrate crystal structure. We analyse the band alignment of the heterostructures and demonstrate a concentration-dependent staggered-to-straddling (typeII-I) band gap transition in PEN/MoSe$_2$. 
\end{abstract}
\vspace{1cm}
\end{@twocolumnfalse}
]

\section{Introduction}
Two-dimensional monolayers of transition metal dichalcogenides (TMDs), namely  MoS$_2$, MoSe$_2$, WS$_2$ and WSe$_2$, readily form van der Waals (vdW) heterostructures with other lattices or molecules \cite{dong2017progress}. The weak vdW interactions allow for easy combinations of heterogeneous materials, potentially giving rise to novel properties \cite{liu2016van, geim2013van}. Monolayer TMDs exhibit favorable optical absorption properties \cite{zhao2015electronic, li2019engineering, kumar2015strong} with many applications, such as photovoltaics \cite{wong2017high}, field-effect transistors \cite{wi2014enhancement, liu2013role} and LED technologies \cite{withers2015light}, all while being thin and flexible enough for wearable, lightweight devices \cite{singh2019flexible}. Additionally, TMD crystals exfoliate to few-layer systems with no dangling bonds, permitting the formation of vdW heterostructures \cite{liu2016van}. Creating these van der Waals heterostructures from TMDs and another semiconducting compound with complimentary electronic properties and band alignment may be a way to improve potential for use in a variety of technological applications, including photovoltaic devices, electrochemical biosensors and chemical catalysts \cite{li2014hemin}. 

Organic compounds are a large family of structures with diverse properties and are often readily synthesised, where the adsorption of organic molecules to TMDs has been shown to modulate their electronic properties \cite{wang2018charge}, in addition to potentially having complimentary properties to the TMD for a given application. One such organic molecule is pentacene (C$_{22}$H$_{14}$). A popular material for combination with TMDs, pentacene boasts strong visible-range spectral absorption \cite{maliakal2004photochemical}, high carrier mobility and photosensitivity \cite{nelson1998temperature, kim2003fabrication}, and as shown in previous works \cite{black2023interaction}, exhibits type-II band alignment when in interface with MoS$_2$, MoSe$_2$, and WS$_2$. Pentacene/MoS$_2$ heterostructures have been shown to exhibit rapid exciton dissociation with recombination lifetimes an order of magnitude longer than those in TMD/TMD heterostructures \cite{bettis2017ultrafast}. Existing literature focuses on thin films of adsorbed pentacene rather than truly 2D structures \cite{habib2020theoretical, markeev2022exciton}, and is mostly limited to adsorption on MoS$_2$. 

\begin{figure}[!ht]
\includegraphics[scale=0.085]{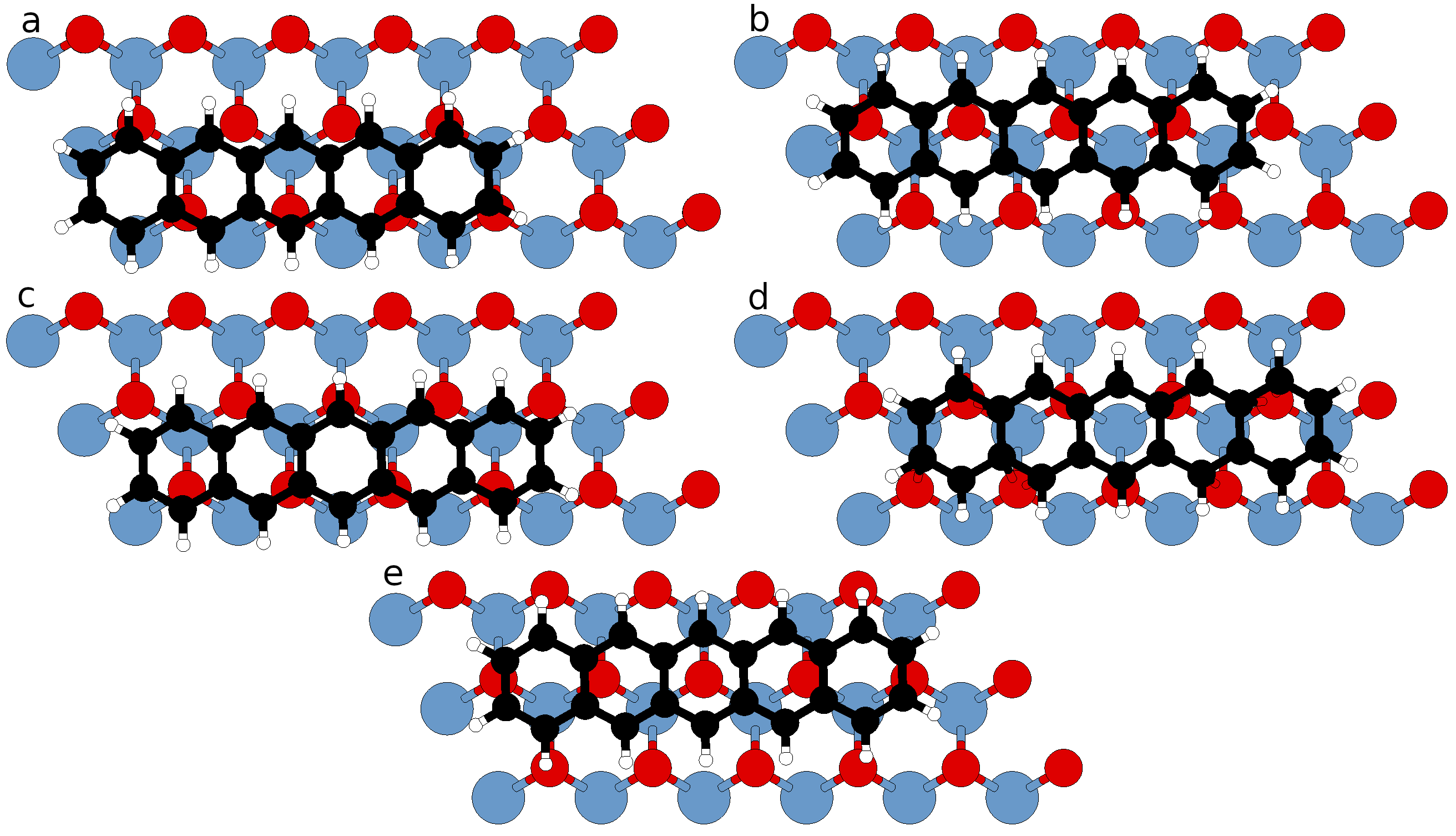}
\caption{Ball and stick representation of investigated adsorption sites of flat-lying, 0\textdegree\ pentacene on 2D monolayer TMD: (a) bridge-A, (b) bridge-B, (c) hollow, (d) top-TM, and (e) top-Ch. Transition metal atoms are shown in blue, and chalcogen atoms in red.}
\label{fig-binding-sites}
\end{figure}

Here systems of pentacene adsorbed upon MoS$_2$, MoSe$_2$, WS$_2$ and WSe$_2$ are investigated, examining the effect of adsorbate concentration and flat-lying rotation with respect to the underlying substrate. Adsorbate concentration was controlled by using one pentacene molecule in two different sizes of supercell; a 6$\times$3 supercell, which corresponds to molecular concentrations of one pentacene molecule per 157 \AA$^2$ (sulphide systems), 170 \AA$^2$ (MoSe$_2$), or 169 \AA$^2$ (WSe$_2$), and a 7$\times$4 supercell, which corresponds to molecular concentrations of one pentacene per 244 \AA$^2$ (sulphide systems), 264 \AA$^2$ (MoSe$_2$), or 263 \AA$^2$ (WSe$_2$).

\section{Computational Details}

\textit{Ab initio} calculations based on density functional theory (DFT) \cite{kohn1965self} were performed on pentacene/TMD heterostructures, with the initial structural parameters and position of the center of mass of pentacene with respect to the underlying substrate determined via geometric optimisation. Pentacene adsorption sites investigated are top-Transition Metal (top-TM), top-Chalcogen (top-Ch), bridge-A, bridge-B and hollow (Figure \ref{fig-binding-sites}).

The calculations were performed using the Quantum Espresso suite \cite{giannozzi2009quantum, giannozzi2017advanced} using the plane wave pseudopotential method within DFT. The Perdew-Burke-Ernzenhof (PBE) exchange-correlation potential approximation \cite{perdew1996generalized}, a generalised gradient approximation (GGA), was used alongside Grimme's DFT-D3 van der Waals force corrections \cite{grimme2010consistent}. Projector-augmented wave (PAW) \cite{blochl1994projector} pseudopotentials \cite{QE-pseudopotentials} were used. A vertical vacuum region of approximately 45 \AA\ maintains repeatd images of the heterostructure isolated from each other in the supercell approach with periodic boundary conditions. For geometric optimization, an energy convergence of $10^{-4}$ Ry was achieved using wavefunction cutoffs of 80, 100, 80 and 120 Ry for MoS$_2$, MoSe$_2$, WS$_2$ and WSe$_2$ respectively (determined after convergence tests). Kinetic energy cut-offs were similarly 320, 400, 320 and 480 Ry, and a Monkhorst-Pack \cite{monkhorst1976special} k-point mesh of 3$\times$6$\times$1 was employed. 
Self-consistent field calculations were performed with the optimized geometry, using the same k-point meshes as in the relaxation calculations. Non-self-consistent field calculations followed, using denser k-point meshes of 6$\times$12$\times$1. Density of states and projected density of states (pDOS) calculations were then performed for the most favorable structures. These calculations were performed with the Gaussian smearing method for Brillouin Zone integration, a simple Gaussian broadening of 0.005 eV and energy grid step of 0.05 eV.
A summary of computational details can be found in Table S1 of the Supplementary Material.

\section{Results}

In this work we investigate both the effects of molecular concentration and rotation on the electronic properties of PEN/TMD heterostructures. We first focused on the molecular concentration, discussed in the first subsection of Results (Section \ref{molecular-concentration}. The effect of rotation is discussed in the following subsection (Section \ref{molecular-rotation}).

\subsection{Effect of molecular concentration}\label{molecular-concentration}

Investigated here is the high concentration regime of pentacene, where we considered a single pentacene molecule adsorbed on a 6$\times$3 supercell of the TMD substrate. The long axis of the pentacene is aligned with the long axis of the underlying substrate crystal (0\textdegree\ of rotation).

In the 6$\times$3 systems minimum molecule-molecule separation on periodic boundary conditions is found to be approximately 3.4 \AA\ for sulphide systems, and 3.7 \AA\ for selenide systems, which have lattice parameters of 3.17 \AA\ for both MoS$_2$ and WS$_2$, and 3.30 \AA\ and 3.29 \AA\ for MoSe$_2$ and WSe$_2$, respectively \cite{black2023interaction}. The same five absorption sites investigated in Ref\cite{black2023interaction} have been revisited here in a higher pentacene concentration regime. These are displayed in Figure \ref{fig-binding-sites} and defined by where on the substrate lattice the central pentacene ring lies over. The central ring of pentacene lies over a bond between a transition metal atom (Mo or W) and a chalcogen atom (S or Se) in the bridge-A and bridge-B sites, over the empty space between within the crystal structure's hexagon in the hollow site, on top of a transition metal atom (Mo or W) in the top-TM site, or over a chalcogen atom (S or Se) in the top-Ch site. Following geometric optimization, top-TM is found to be the most favorable binding site for MoS$_2$, MoSe$_2$ finds bridge-B to be most favorable, WS$_2$ finds top-Ch to be most favorable, and WSe$_2$ finds bridge-A to be most favorable (Table \ref{tab-adsorption-energy-6x3}). In MoS$_2$, top-TM is 48 meV more favorable than the next-most favorable site (bridge-B), and the difference between most and least favorable sites being 143 meV. MoSe$_2$ shows a smaller difference between binding site favorability, the difference between most and next-most favorable site being only 5 meV, with a difference between most and least favorable being 89 meV. This indicates high molecular mobility in MoSe$_2$, as suggested experimentally in \cite{kachel2021engineering} for the case of MoS$_2$, and is very similar to the energy differences between binding sites found amongst all TMDs in previous work \cite{black2023interaction}. WS$_2$ is mobile between the preferred top-Ch site and the next most favorable site of bridge-B, with a difference of only 12 meV. A difference of 87 meV separates the most and least favorable sites in WS$_2$, implying good mobility across all sites, as in MoSe$_2$. WSe$_2$ shows a difference of 137 meV between most and next-most favorable sites and 188 meV between most and least favorable sites. The pentacene molecule lies flat in all systems, with minimal tilting and bending. Binding distances for the most favorable adsorption sites, determined by measuring separation between the chalcogenide top layer (that closest to the pentacene molecule) of the substrate surface and the centre of mass of the pentacene molecule, were 3.30 \AA, 3.42 \AA, 3.32 \AA\ and 3.48 \AA\ for MoS$_2$, MoSe$_2$, WS$_2$ and WSe$_2$ respectively. These values are similar in magnitude and nature to those found previously in the lower concentration (7$\times$4 supercell) systems, with larger binding distances in Se systems, notably in spite of the larger adsorption energy of the WSe$_2$ system in particular (Table \ref{tab-adsorption-energy-6x3}). This is again thought to be explained by the larger vdW radius of a selenium atom compared to a sulphur atom (1.90 \AA\ and 1.73 \AA, respectively \cite{batsanov2001van}). 

Additionally, the top-TM site of MoS$_2$ is not stable, and shifts under relaxation to between top-TM and hollow sites. This intermediate site between top-TM and hollow (Figure S1 of the Supplementary Material) is, however, more favorable than the other sites investigated for this system. When rotating the pentacene molecule on 6$\times$3 MoS$_2$, both the shifted site and the 'true' top-TM site (where the shift towards hollow was not permitted) were investigated. The 'true' top-TM system was less favorable than the intermediate site when rotated and fully relaxed, and so was not investigated further. As the 'true' top-TM site is not entertained going forward, the intermediate binding site shall be referred to simply as top-TM, as the pentacene was over this site prior to initial relaxation. 

\begin{table*}[!h]
    \centering
    \caption{Adsorption energies, in eV, following geometry optimization of the 6$\times$3 PEN/TMD heterostructures. Most favorable binding sites are highlighted in bold, and binding site geometries are displayed in Figure \ref{fig-binding-sites}. $^{\dagger}$The initial binding site of top-TM is unstable, and shift to a stable intermediate position (Figure S1 of the Supplementary Material).}
    \begin{tabular}{cccccc}
        \hline
         TMD & Bridge-A & Bridge-B & Hollow & Top-TM & Top-Ch \\
         \hline
         MoS$_2$ & -1.318 & -1.413 & -1.363 & \textbf{-1.461}$^{\dagger}$ & -1.407 \\
         MoSe$_2$ & -1.348 & \textbf{-1.437} & -1.387 & -1.377 & -1.432 \\
         WS$_2$ & -1.367 & -1.443 & -1.393 & -1.392 & \textbf{-1.454} \\
         WSe$_2$ & \textbf{-1.606} & -1.469 & -1.420 & -1.419 & -1.466 \\
         \hline 
         \hline
    \end{tabular}
    \label{tab-adsorption-energy-6x3}
\end{table*}

Adsorption energies ($E_{ads}$) of pentacene/TMD systems, displayed in Table \ref{tab-adsorption-energy-6x3}, were calculated as the difference between the combined system's energy ($E_{PEN/TMD}$) and the sum of the energies of the isolated systems with independently relaxed geometry ($E_{TMD}^{relax}$ and $E_{PEN}^{iso-relax}$) as in Equation 1.

\begin{equation}
    E_{ads} = E_{PEN/TMD} - E_{TMD}^{relax} - E_{PEN}^{iso-relax}
\end{equation}

A comparison of the structural properties between 0\textdegree\ pentacene adsorbed on 6$\times$3 TMD and previous work on 7$\times$4 TMD \cite{black2023interaction} is provided in Table \ref{tab-6x3vs7x4-structure}. 7$\times$4 systems consistently preferred a top-Ch binding site, while there was variation within the 6$\times$3 systems. The 7$\times$4 systems demonstrated consistently lower differences in adsorption energy between binding sites, between 2 meV and 6 meV higher energy than the most favorable top-Ch site for the next-most favorable (bridge-B in all 7$\times$4 systems) site, and between 24 meV and 83 meV  higher energy between the other sites \cite{black2023interaction}. In contrast, the 6$\times$3 systems demonstrate a much larger range of differences (between 5 meV and 188 meV), with overall larger energy differences between sites (the 5 meV being followed by the next smallest energy difference of 48 meV), implying a much more restricted molecular mobility between binding sites than found in the 7$\times$4 systems. Adsorption energies for favorable 7$\times$4 TMDs increased in magnitude with increasing substrate mass, with MoSe$_2$ and WS$_2$ being similar, and the 6$\times$3 systems did not follow this pattern. However; as they favor different binding sites, a direct comparison of adsorption energies is not expected to provide the same pattern.

Separation between pentacene and TMD varied in the same manner for both concentrations (see Table \ref{tab-6x3vs7x4-structure}), with the sulphide systems showing binding at closer range, with the 7$\times$4 systems binding tighter than their 6$\times$3 counterparts, in particular for WSe$_2$. A similar finding has been previously reported for pentacene adsorbed on metal Ag(111) surface, where binding distances were found to increase with molecular concentration due to competing effects of molecule-molecule and molecule-substrate interactions \cite{duhm2013pentacene}. 

\begin{figure*}
    \centering
    \includegraphics[scale=0.3]{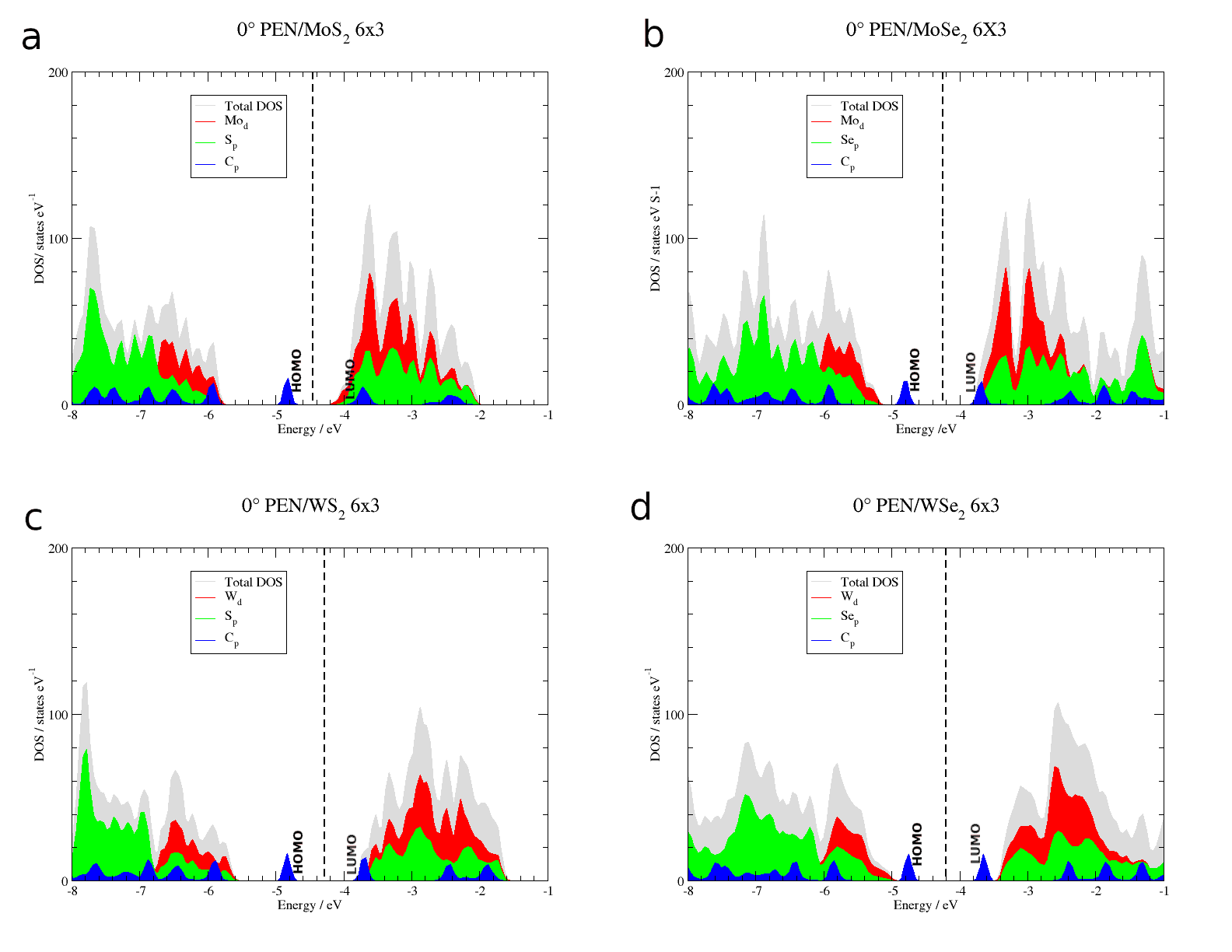}
    \caption{Total and projected density of states (DOS) of PEN/TMD systems in the 6$\times$3 supercell: PEN/MoS$_2$ (a), PEN/MoSe$_2$ (b), PEN/WS$_2$ (c), and PEN/WSe$_2$ (d). Fermi energy is denoted by a vertical dashed line, and pentacene's HOMO and LUMO are labelled.}
    \label{fig-pDOS-unrotated-6x3}
\end{figure*}

Density of States (DOS) was investigated for the most favorable binding sites of each TMD system in the 6$\times$3 supercell, in all cases demonstrating that pentacene's highest occupied molecular orbital (HOMO), contributed by carbon p-orbitals, is located within the TMD band gap (Figure \ref{fig-pDOS-unrotated-6x3}). This state is closer to the TMD's valence band maximum (VBM) in the selenide systems than it is in the sulphide systems, explaining in part why there is stronger interaction between pentacene and WSe$_2$ than in WS$_2$. This difference in proximity is not as marked in the molybdenum systems, and so the same difference in adsorption energy is not seen between MoSe$_2$ and MoS$_2$.

\begin{table*}
\caption{Structural comparison of 0\textdegree\ pentacene adsorbed on TMDs for 6$\times$3 (higher molecular concentration) and 7$\times$4 (lower molecular concentration) supercells. 7$\times$4 parameters are from Ref\cite{black2023interaction}.}
\begin{tabular}{ccccc}
    \hline
     TMD & MoS$_2$ & MoSe$_2$ & WS$_2$ & WSe$_2$ \\
     \hline
     6$\times$3 & & & & \\
     Binding Site & Top-TM & Bridge-B & Top-Ch & Bridge-A \\
     Adsorption Energy/eV & -1.46 & -1.44 & -1.45 & -1.61 \\
     Binding Distance/\AA & 3.30 & 3.42 & 3.32 & 3.48 \\
     Minimum molecule-molecule distance/\AA & 3.4 & 3.7 & 3.4 & 3.7 \\
     \hline
     7$\times$4 & & & & \\
     Binding Site & Top-Ch & Top-Ch & Top-Ch & Top-Ch \\
     Adsorption Energy/eV & -1.39 & -1.42 & -1.43 & -1.46 \\
     Binding Distance/\AA & 3.31 & 3.40 & 3.30 & 3.38 \\
     Minimum molecule-molecule distance/\AA & 6.2 & 6.5 & 6.2 & 6.5 \\
     \hline
     \hline
\end{tabular}
\label{tab-6x3vs7x4-structure}
\end{table*}

Pentacene's lowest unoccupied molecular orbital (LUMO) is located above the conduction band minimum (CBM) of sulphide systems (although this is a close distinction in WS$_2$, see Figures \ref{fig-pDOS-unrotated-6x3}a and \ref{fig-pDOS-unrotated-6x3}c), indicating a staggered band gap, and so that these systems are type-II heterostructures. Selenide systems (Figure \ref{fig-pDOS-unrotated-6x3}b and \ref{fig-pDOS-unrotated-6x3}d), however, show a type-I band alignment, with pentacene's LUMO of lower energy than the TMD's CBM, with both of pentacene's HOMO and LUMO within the TMD's band gap in the case of the selenide systems, seen as carbon p-orbital states in Figure \ref{fig-pDOS-unrotated-6x3}b and \ref{fig-pDOS-unrotated-6x3}d. In PEN/WS$_2$ (Figure \ref{fig-pDOS-unrotated-6x3}c), pentacene's LUMO is very close to the TMD's CBM, but does display type-II alignment.

A comparison between the DOS of the two concentration regimes (6$\times$3 and 7$\times$4 supercells \cite{black2023interaction}) is shown in Figure S6 of the Supplementary Material. The DOS of the two concentration regimes are strikingly similar, both sets showing an inter-gap state contributed by the pentacene's carbon p-orbital, closer to the TMD's VBM in selenide systems than in sulphide, and creating a type-II band alignment in sulphide systems in both concentrations investigated, type-I band alignment in WSe$_2$ structures in both concentrations, and a concentration dependency in PEN/MoSe$_2$. The Fermi energies are similar, with PEN/MoS$_2$ decreasing from -4.42eV to -4.45eV with increased pentacene concentration, PEN/MoSe$_2$ decreasing from -4.21eV to -4.25eV, and PEN/WSe$_2$ from -4.17eV to -4.20eV, but PEN/WS$_2$ increasing from -4.28eV to -4.26eV. The positions of pentacene's HOMO and LUMO are also similar, with changes in LUMO energy due to increased pentacene concentration of -0.08eV (PEN/MoS$_2$), -0.04eV (PEN/MoSe$_2$), -0.08eV (PEN/WS$_2$) an -0.09eV (PEN/WSe$_2$). Despite having the smallest change in LUMO, PEN/MoSe$_2$ undergoes a band alignment transition due to the proximity of pentacene's LUMO to the CBM in the 7$\times$4 supercell, with the small change in LUMO from increasing concentration being enough to cause the transition. HOMO and LUMO energies of pentacene in both higher (6$\times$3) and lower (7$\times$4) concentrations can be found in Table S3 of the Supplementary Material. 

\subsection{Effect of adsorbate rotation}\label{molecular-rotation}

\begin{table*}
\centering
\caption{Adsorption energies in eV following relaxation of rotated PEN/TMD systems in the 6$\times$3 supercell, starting from the most energetically favorable binding site (Tables \ref{tab-adsorption-energy-6x3} and \ref{tab-6x3vs7x4-structure}). $^{\dagger}$90\textdegree is the starting angle, but is unstable and shifts to 79\textdegree\ in MoS$_{2}$, 70\textdegree\ in WS$_2$, or 74\textdegree in selenide systems.}
\begin{tabular}{ccccc}
        \hline
         TMD & 0\textdegree & 30\textdegree & 60\textdegree & 90\textdegree$^{\dagger}$ \\
         \hline
         MoS$_2$ & \textbf{-1.461} & -1.371 & -1.299 & -1.302 \\
         MoSe$_2$ & \textbf{-1.437} & -1.385 & -1.356 & -1.405 \\ 
         WS$_2$ & -1.455 & -1.427 & \textbf{-1.461} & -1.413 \\
         WSe$_2$ & -1.606 & -1.637 & \textbf{-1.682} & -1.667 \\
         \hline
         \hline
\end{tabular}
\label{tab-adsorption-energy-rotated-6x3}
\end{table*}

\begin{figure}[!ht]
\includegraphics[scale=0.1]{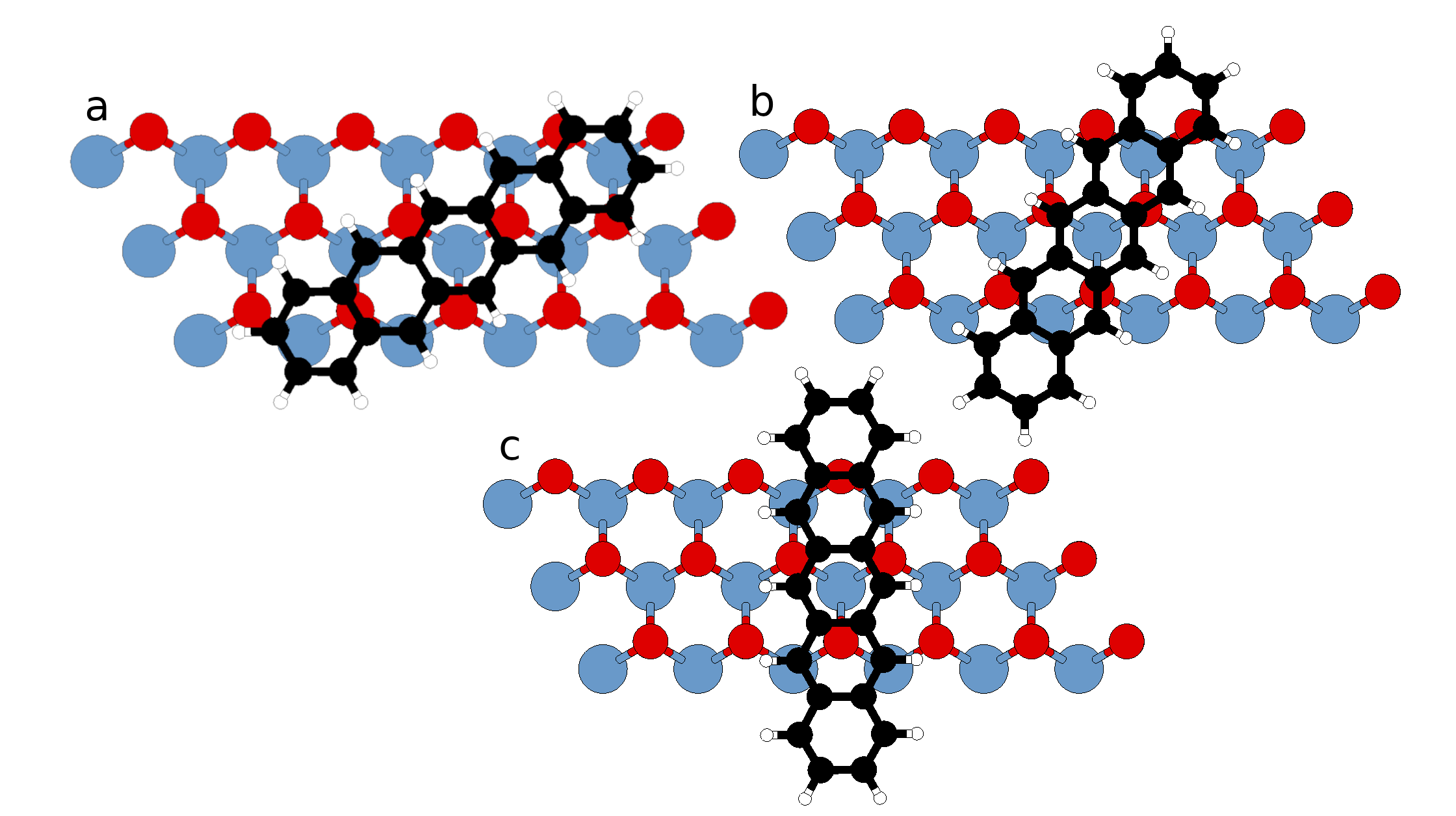}
\caption{Ball and stick representation of rotated pentacene adsorbed on 2D monolayer TMD: (a) 30\textdegree, (b) 60\textdegree, and (c) 90\textdegree.}
\label{fig-rotated-6x3}
\end{figure}

The pentacene molecule was then rotated about its center of mass in the plane of the two-dimensional substrate lattice (pentacene remains flat-lying and maintains previously determined inter-layer separation) over the previously determined most favorable binding site for PEN/TMD systems. Rotation was with respect to the centre of mass of pentacene, with counterclockwise rotation angles of 30\textdegree, 60\textdegree\ and 90\textdegree\ being investigated. 0\textdegree\ is defined as aligned with the long axis of the underlying substrate within the supercell. The rotated structures were then geometrically optimised, and the electronic properties of the resulting systems investigated. 7$\times$4 supercell systems were likewise rotated and compared, with starting geometry from Ref\cite{black2023interaction}. The initial MoS$_2$ systems following the rotation of pentacene, but before geometric optimization, are shown in Figure \ref{fig-rotated-6x3}; these are qualitatively representative of the other TMD systems. 

\subsubsection{Higher molecular concentration: one pentacene molecule per 6$\times$3 supercell}

As displayed in Table \ref{tab-adsorption-energy-rotated-6x3}, when rotating pentacene in the high concentration regime for favorable binding sites, PEN/MoS$_2$ favored 0\textdegree\ pentacene, as did MoSe$_2$. The tungsten systems, however, were more energetically favorable after undergoing rotation: both WS$_2$ and WSe$_2$ favored a 60\textdegree\ rotation, which remains stable under relaxation. None of the high pentacene concentration systems remained stable under relaxation when starting at 90\textdegree rotated pentacene; MoS$_2$ relaxed to an angle of 79\textdegree, WS$_2$ to 70\textdegree, and the selenide systems to 74\textdegree. 30\textdegree\ and 60\textdegree\ starting angles remained stable under relaxation in all systems. Minimum molecule-molecule separations of the favorable rotated systems were 3.1 \AA\ and 3.4 \AA\ for WS$_2$ and WSe$_2$ respectively. Table \ref{tab-6x3-favorability} summarises the favorability of the rotated high concentration regime.
\newline

\begin{table*}
\centering
\caption{Summary of the structural properties of PEN/TMD with 6$\times$3 supercells, considering rotation of the adsorbate.}
\begin{tabular}{ccccc}
        \hline
         TMD & Binding Site & Rotation & Adsorption Energy/eV & Binding Distance/\AA\ \\
         \hline
         MoS$_2$ & Top-TM & 0\textdegree & -1.461 & 3.30\\
         MoSe$_2$ & Bridge-B & 0\textdegree & -1.437 & 3.42 \\
         WS$_2$ & Top-Ch & 60\textdegree & -1.461 & 3.32 \\
         WSe2 & Bridge-A & 60\textdegree & -1.682 & 3.46 \\
         \hline
         \hline
\end{tabular}
\label{tab-6x3-favorability}
\end{table*}

Unfavorable rotation angles for PEN/MoS$_2$ are between 90 meV and 160 meV less energetically favorable, implying a lack of mobility in angle of rotation, similar to the immobile binding sites in 6$\times$3 PEN/TMD systems, and in contrast to those in the 7$\times$4 systems, while PEN/MoSe$_2$ demonstrates higher energies of 32 meV to 81 meV for rotated systems. PEN/WS$_2$ has higher energies of 6 meV to 48 meV of the less favorable rotations than the most favorable 60\textdegree, but the least favorable investigated is the 70\textdegree\ rotation, and next-most favorable is the 0\textdegree\ system. WSe$_2$ demonstrates a difference between the most favorable 60\textdegree\ and next-most favorable 74\textdegree\ of only 15 meV, and a difference of 76 meV between most and least favorable rotations. This implies some mobility between the 60\textdegree\ and 74\textdegree\ rotations, with the starting angle of 90\textdegree\ to 74\textdegree\ or 79\textdegree\ supporting the concept in these systems, except for WS$_2$ (although the difference is only 48 meV, not much higher than MoSe$_2$'s difference to its next most favorable rotation).

Minimum molecule-molecule separation is smaller in the rotated systems, but molecule-molecule interaction is not a significant enough contributor to adsorption energy to prevent the tungsten systems from being more favorable in a rotated geometry. Binding distances were 3.32 \AA\ and 3.46 \AA\ for WS$_2$ and WSe$_2$, respectively (see Table \ref{tab-6x3-favorability}), demonstrating no change in distance in the WS$_2$ system compared to its 0\textdegree\ geometry, and only a small change in the WSe$_2$ system. 

By calculating the adsorption energy using isolated systems with independently relaxed geometries, contributions from molecule-substrate and molecule-molecule interactions, as well as molecule and substrate deformation can be calculated as previously done in Ref\cite{black2023interaction}, and are summarised in Table \ref{tab-6x3-rotated-adsorption-energy-contributions}.
\newline

\begin{table*}
\centering
\caption{Contributions towards the adsorption energy of favorable rotated 6$\times$3 systems of PEN/TMD from molecule-molecule and molecule-substrate interactions, and molecule and substrate deformation, in eV. Angles of rotation are given in parenthesis. Details of these calculations can be found in the Supplementary Material}
\begin{tabular}{ccccc}
        \hline
        TMD & MoS$_2$ (0\textdegree) & MoSe$_2$ (0\textdegree) & WS$_2$ (60\textdegree) & WSe$_2$ (60\textdegree) \\
        \hline
        molecule-molecule interaction & -0.0592 & -0.0375 & -0.0651 & -0.0381 \\
        molecule-substrate interaction & -1.398 & -1.400 & -1.393 & -1.422 \\
        molecule deformation & -0.0088 & -0.0053 & -0.0093 & -0.0072 \\
        substrate deformation & 0.0016 & 0.0025 & 0.0032 & -0.2180 \\ 
        \hline
        \hline
\end{tabular}
\label{tab-6x3-rotated-adsorption-energy-contributions}
\end{table*}

The largest contribution to adsorption energy is due to the molecule-substrate interaction, with other contributions being orders of magnitude smaller. This would be expected as the pentacene molecule does not deform during relaxation. Molecule-molecule interaction is one order of magnitude higher than that calculated for 7$\times$4 supercells in Ref\cite{black2023interaction}, which is a consequence of the smaller separation between the molecules across periodic boundary conditions. 

A comparison of the DOS of 0\textdegree\ and favorable rotated systems shows no large deviation, with band alignment unchanged, comparable Fermi energies, and similar projected DOS topography, as shown in Figure S3 of the Supplementary Material.

\subsubsection{Lower molecular concentration: one pentacene molecule per 7$\times$4 supercell}

In the low concentration pentacene regime (7$\times$4 supercell of TMD crystal) all systems were previously found to be most favorable with pentacene at the top-Ch site \cite{black2023interaction}. All four systems were found to be more energetically favorable when rotated by 60\textdegree\ about the centre of mass from the original 0\textdegree\ orientation (Table \ref{tab-adsorption-energy-rotated-7x4}), in contrast to the high concentration regime where there was variability between the TMDs (similar to binding site). All systems were stable about their starting angles, and remained flat-lying without significant bending or tilting. The use of a 7$\times$4 TMD crystal supercell yields a minimum molecule-molecule separation of approximately 6.2 \AA\ for 0\textdegree\ sulphide systems, and 6.5 \AA\ for 0\textdegree\ selenide systems. When rotated, the minimum separation decreases to approximately 5.2 \AA\ (6.0 \AA) for 30\textdegree\ rotation, 5.9 \AA\ (6.3 \AA) for 60\textdegree\ rotation, and 1.9 \AA\ (2.5 \AA) for 90\textdegree\ rotation on sulphide (selenide) systems. Similar to previous observation in the 0\textdegree\ systems, we notice only a small number of carbon atoms located over TMD atoms, with many over the hollow site, reducing steric repulsion in the heterostructure with rotated pentacene. This is a trend noticed in both the sulphide and the selenide systems. Binding distances of these 60\textdegree\ systems are 3.31 \AA, 3.40 \AA, 3.30 \AA, and 3.39 \AA\ for MoS$_2$, MoSe$_2$, WS$_2$ and WSe$_2$, respectively. These are exceptionally similar to the unrotated systems \cite{black2023interaction}, following the same pattern of more closely bound for molybdenum compared to tungsten systems (as well as having lower adsorption energies), but greater binding distances for selenide compared to sulphide systems.

As can be seen in Table \ref{tab-adsorption-energy-rotated-7x4}, the difference in adsorption energies between the 60\textdegree\ rotations and other angles were between 1 meV and 30 meV for MoS$_2$, 1 meV and 38 meV in MoS2$_2$, and 2 meV and 38 meV in WS$_2$. WSe$_2$ does not share these similar energy discrepancies, with a range between 11 meV and 355 meV. All TMDs except WSe$_2$ showed only a slight energetic preference for 60\textdegree\ over 0\textdegree\ pentacene, with the least favorable rotation being 30\textdegree. WSe$_2$ instead showed a large energy difference between 0\textdegree\ and 60\textdegree\ pentacene, with 30\textdegree\ rotation being comparably less favorable than with other TMDs, and 90\textdegree\ being the next-most favorable rotation after 60\textdegree. Pentacene adsorbed on TMDs, other than WSe$_2$, may therefore be mobile within its z-axis rotational degree of freedom around the 0\textdegree\ angles, with reduced mobility as it approaches 30\textdegree, which is an unfavorable angle, creating a local minima of adsorption energy. Beyond this energy barrier lies a minima of lower energy than around 0\textdegree\, somewhere around 60\textdegree. WSe$_2$ exhibits different behaviour, with a very unfavorable position in 0\textdegree\, becoming more favorable as rotation angle increases away from this point. 

\begin{table*}
\centering
\caption{Adsorption energies in eV following relaxation of rotated PEN/TMD systems in the 7$\times$4 supercells, starting from the most energetically favorable binding site (top-Ch in all cases \cite{black2023interaction}).}
\begin{tabular}{ccccc}
        \hline
         TMD & 0\textdegree & 30\textdegree & 60\textdegree & 90\textdegree \\
         \hline
         MoS$_2$ & -1.389 & -1.361 & \textcolor{red}{-1.390} & -1.370 \\
         MoSe$_2$ & -1.424 & -1.386 & \textcolor{red}{-1.425} & -1.422 \\
         WS$_2$ & -1.434 & -1.399 & \textcolor{red}{-1.436} & -1.407 \\
         WSe$_2$ & -1.458 & -1.768 & \textcolor{red}{-1.813} & -1.802 \\
         \hline
         \hline
\end{tabular}
\label{tab-adsorption-energy-rotated-7x4}
\end{table*}

In the same manner as for the 6$\times$3 TMD supercell systems, the contributions to adsorption energy from molecule-molecule interaction, molecule-substrate interaction and deformation of the molecule and of the substrate can be calculated and are displayed in Table \ref{tab-7x4-rotated-adsorption-energy-contributions}. Again it is found that molecule-substrate interaction is the largest contributor, but in the low concentration regime other sources are orders of magnitude smaller only in molybdenum systems. In both tungsten systems, substrate deformation is responsible for a relatively large amount of the adsorption energy, and in the case of WS$_2$, \textit{opposes} heterostructure formation. In the unrotated systems, this opposition is found in substrate deformation for all systems \textit{except} WS$_2$, but it is of much smaller magnitude \cite{black2023interaction}. As would be expected, the larger supercell (leading to greater separation of molecules when compared to the 6$\times$3 regime) results in a molecule-molecule interaction contribution much less significant than in the high concentration pentacene regime. No significant difference was observed when compared to 0\textdegree\ rotation \cite{black2023interaction}.

\begin{table*}
\centering
\caption{Contributions towards the adsorption energy of favorable 7$\times$4 systems following rotation of pentacene from molecule-molecule and molecule-substrate interactions, and molecule and substrate deformation, in eV. The most favorable angle of rotation for each TMD is given in parenthesis.}
\begin{tabular}{ccccc}
        \hline
        TMD & MoS$_2$ (60\textdegree) & MoSe$_2$ (60\textdegree) & WS$_2$ (60\textdegree) & WSe$_2$ (60\textdegree) \\
        \hline
        molecule-molecule interaction & -0.0048 & -0.0034 & -0.0048 & -0.0035 \\
        molecule-substrate interaction & -1.393 & -1.422 & -1.439 & -1.461 \\
        molecule deformation & -0.0067 & -0.0100 & -0.0064 & -0.0094 \\
        substrate deformation & 0.0060 & 0.0011 & 0.10017 & -0.3456 \\
        \hline
        \hline
\end{tabular}
\label{tab-7x4-rotated-adsorption-energy-contributions}
\end{table*}

A comparison of the DOS of 0\textdegree\ heterostructures and favorable rotated systems again shows no marked change with unchanged band alignment, and comparable Fermi energies and projected DOS contributions (Figure S5 of the Supplementary Material).

\subsection{A comparison of favorably rotated systems}

Table \ref{tab-6x3vs7x4-rotated-structure} is a comparison of structural properties between the favorable 6$\times$3 and 7$\times$4 TMD supercell systems, after rotation. The 7$\times$4 systems were consistent in their preferred binding site, while the 6$\times$3 systems were not. It is noted that most systems prefer 60\textdegree\ rotation, and it is the systems with lowest mass per supercell that does not prefer 60\textdegree; as the total substrate mass per supercell increases, and so does electron density (8.51 e$^{-}$/\AA$^2$, 11.68 e$^{-}$/\AA$^2$, 12.16 e$^{-}$/\AA$^2$ and 15.13 e$^{-}$/\AA$^2$ for MoS$_2$, MoSe$_2$, WS$_2$ and WSe$_2$, respectively, the same for both 6$\times$3 and 7$\times$4 systems). In the 7$\times$4 systems, and $6\times$3 selenide systems, the preferred angle is 60\textdegree. The $6\times$3 systems exhibit differences in adsorption energies between the most favorable and the next-most favorable rotations of between 90 meV, 32 meV, 6 meV and 15 meV for MoS$_2$, MoSe$_2$, WS$_2$ and WSe$_2$, respectively. The 7$\times$4 systems instead are much more mobile in their pentacene's z-axis rotational freedom, as previously discussed, with differences of 1 meV for both molybdenum systems, 2 meV for WS$_2$, and a slightly larger 11 meV for WSe$_2$.

Binding distances followed the pattern shown by 0\textdegree\ systems, with very similar separation between layers compared to both concentration regime's unrotated counterparts. The 6$\times$3 tungsten systems, that preferred an angle other than 0\textdegree, adsorb at a greater distance from the substrate than the 7$\times$4 tungsten systems, but the molybdenum systems are similar between concentration regimes. 

\begin{table*}
    \centering
    \caption{Structural comparison of favorable systems of pentacene adsorbed on TMDs, considering rotation of the pentacene molecule.}
    \begin{tabular}{ccccc}
        \hline
         TMD & MoS$_2$ & MoSe$_2$ & WS$_2$ & WSe$_2$ \\
         \hline
         6$\times$3 & & & & \\
         Angle of Rotation & 0\textdegree & 0\textdegree & 60\textdegree & 60\textdegree \\
         Adsorption Energy/eV & -1.46 & -1.44 & -1.44 & -1.68 \\
         Binding Distance/\AA & 3.30 & 3.42 & 3.32 & 3.46 \\
         Minimum molecule-molecule distance/\AA & 3.4 & 3.7 & 3.1 & 3.4 \\
         \hline
         7$\times$4 & & & & \\
         Angle of Rotation & 60\textdegree & 60\textdegree & 60\textdegree & 60\textdegree \\
         Adsorption Energy/eV & -1.39 & -1.42 & -1.44 & -1.81 \\
         Binding Distance/\AA & 3.31 & 3.40 & 3.30 & 3.39 \\
         Minimum molecule-molecule distance / \AA & 5.9 & 6.3 & 5.9 & 6.3 \\
         \hline
         \hline
    \end{tabular}
    \label{tab-6x3vs7x4-rotated-structure}
\end{table*}

Density of states calculations show no changes to the character of the heterojunctions upon decreasing molecular concentration on MoS$_2$, WS$_2$ or WSe$_2$ substrates, with PEN/WSe$_2$ remaining a type-I heterojunction and the other sulphide systems exhibiting type-II band alignment. The PEN/MoSe$_2$ heterojunction, however, undergoes a transition form a type-II (staggered) band alignment in the lower concentration regime (7$\times$4 supercell) to a type-I (straddled) band alignement with increased molecular concentration (6$\times$3 supercell). All systems maintain their intergap state contributed by pentacene's carbon p-orbital and the Fermi energies are negligibly changed, although it is noted that changing the concentration of pentacene has a larger effect on the Fermi energy than rotating the adsorbate alone.

\begin{figure*}
    \centering
    \includegraphics[scale=0.3]{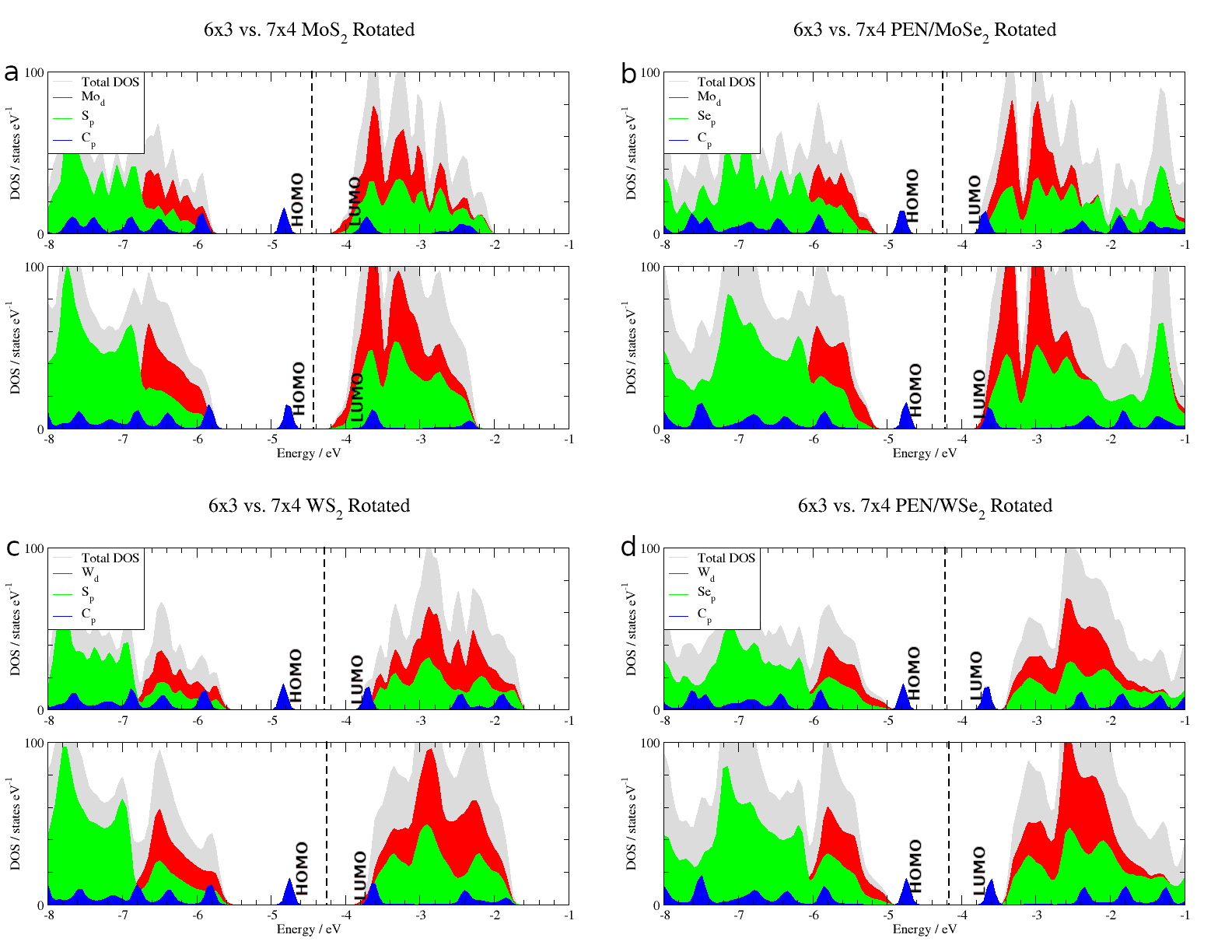}
    \caption{Total and projected density of states of energetically favorable rotated pentacene on 6$\times$3 (top panels) and 7$\times$4 (bottom panels): PEN/MoS$_2$ (a) with 0\textdegree pentacene (top panel) and 60\textdegree\ pentacene (bottom panel), PEN/MoSe$_2$ (b) with 0\textdegree\ (top panel) and 60\textdegree\ (bottom panel) pentacene, PEN/WS$_2$ (c) with 60\textdegree\ (top panel) and 60\textdegree\ pentacene (bottom panel), and PEN/WSe$_2$ (d) with 0\textdegree\ (top panel) and 60\textdegree\ (bottom panel) pentacene.}
    \label{fig-pDOS-comparison-rot}
\end{figure*}

\section{Conclusions}
In summary, adsorbed pentacene of two different concentrations and four angles of rotation with respect to the substrate's long axis have been investigated on the 2D monolayer TMD substrates of MoS$_2$, MoSe$_2$, WS$_2$ and WSe$_2$. Pentacene lies flat in all systems, and binds more strongly when in the lower concentration. High concentration pentacene yields variation in favorable adsorption site and angle of rotation between the investigated TMDs, with mobility between adsorption sites being generally reduced when compared to the lower concentration regime, whereas low concentration pentacene adsorbs most favorably in the same site and at the same angle across all investigated TMDs. Band alignment of the PEN/TMD heterostructures is robust to changes in the angle of pentacene and the change in concentration investigated here (but band tuning was observed), with the exception of MoSe$_2$. PEN/MoSe$_2$ undergoes a band alignment transition between the low and high pentacene concentration regimes, from a type-II to a type-I.

\section{Acknowledgements}
This work was partially funded by the Deutsche Forschungsgemeinschaft (DFG, German Research Foundation), Project 406901005. Calculations were performed using the Cirrus UK National Tier-2 HPC Service at EPCC (http://www.cirrus.ac.uk) funded by the University of Edinburgh and EPSRC (EP/P020267/1).

\section{Data availability statement}

The data that support the findings of this study are openly available at Keele Data Repository: \textit{URL to be provided soon}.

\bibliography{references.bib}
\bibliographystyle{ieeetr}
\end{document}

\maketitle
\renewcommand{\thetable}{S\arabic{table}}
\renewcommand{\thefigure}{S\arabic{figure}}
\section{Computational Details}
Table \ref{supp-tab-comp-details} summarises the computational details of the density functional theory calculations performed in this work.

\begin{table}[H]
\centering
\caption{Computational details of the calculations performed in this work.}
\begin{tabular}{cc}
\hline
\hline
     code & Quantum ESPRESSO \cite{giannozzi2009quantum, giannozzi2017advanced} \\
     theory & DFT \cite{kohn1965self} \\
     exchange-correlation functional & GGA-PBE \cite{perdew1996generalized} \\
     van der Waals method & DFT-D3 \cite{grimme2010consistent} \\
     pseudopotentials & PAW \cite{blochl1994projector, QE-pseudopotentials} \\
     & Mo.pbe-spn-kjpaw\_psl.1.0.0.UPF \\
     & W.pbe-spn-kjpaw\_psl.1.0.0.UPF \\
     & S.pbe-p-kjpaw\_psl.1.0.0.UPF \\
     & Se.pbe-dn-kjpaw\_psl.1.0.0.UPF \\
     & C.pbe-n-kjpaw\_psl.1.0.0.UPF \\
     & H.pbe-kjpaw\_psl.1.0.0.UPF \\
     wavefunction energy cutoff & MoS$_2$: 80 Ry, MoSe$_2$: 100 Ry, WS$_2$: 80 Ry, WSe$_2$: 120 Ry \\
     supercell & high molecular concentration: \\
     & 6$\times$3 TMD supercell, 45 \AA\ vacuum region \\
     & low molecular concentration: \\
     & 7$\times$4 TMD supercell, 45 \AA\ vacuum region \\
     lattice parameters & MoS$_2$: 3.17 \AA, MoSe$_2$: 3.30 \AA, WS$_2$: 3.17 \AA, WSe$_2$: 3.29 \AA \\
     k-point grid & 3$\times$6$\times$1 Monkhorst-Pack grid \cite{monkhorst1976special} \\
     \hline
     \hline
\end{tabular}
\label{supp-tab-comp-details}
\end{table}

\section{Structure}

\begin{figure}[H]
    \centering
    \includegraphics[scale=0.15]{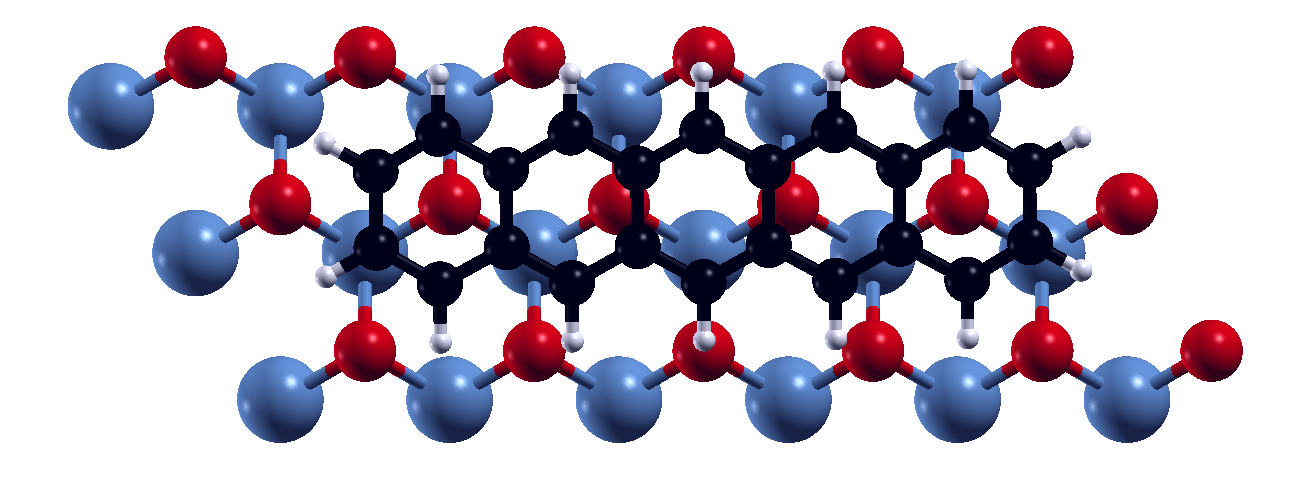}
    \caption{The intermediate adsorption site between top-TM and hollow that the 6$\times$3 PEN/MoS$_2$ system relaxed to following a starting position of top-TM. This system was most energetically favorable in this site.}
    \label{supp-fig-shifted-MoS2-6x3}
\end{figure}

During relaxation of the 6$\times$3 binding site geometry, the top-TM site for MoS$_2$ was unstable, and shifted towards the hollow site. The geometry following relaxation is shown in Figure \ref{supp-fig-shifted-MoS2-6x3}. Other binding sites were stable.

\section{Adsorption energies}

Contributions to the adsorption energy, $E_{ads}$, due to molecule-molecule interaction, $E^{int}_{mol}$, deformation of the molecule, $E^{def}_{mol}$, molecule-substrate interaction, $E^{int}_{mol-TMD}$ and substrate deformation, $E^{def}_{TMD}$ were calculated both in this work and in Reference \cite{black2023interaction} usign the following;

\begin{equation}
    E^{int}_{mol} = E^{scell-adsgeom}_{mol} - E^{iso-adsgeom}_{mol}
\end{equation}
\begin{equation}
    E^{def}_{mol} = E^{iso-adsgeom}_{mol} - E^{iso-relax}_{mol}
\end{equation}
\begin{equation}
    E^{int}_{mol-TMD} = E_{PEN/TMD} - E^{adsgeom}_{TMD} - E^{scell-adsgeom}_{mol}
\end{equation}
\begin{equation}
    E^{def}_{TMD} = E^{adsgeom}_{TMD} - E^{relax}_{TMD}
\end{equation}

where $E^{iso-relax}_{mol}$ is the total energy of isolated and relaxed pentacene in a cubic supercell of side 48 \AA, $E^{iso-adsgeom}_{mol}$ is the total energy of isolated pentacene with adsorbed geometry in the same supercell, and pentacene with adsorbed geometry in the supercell of the heterostructure (either 6$\times$3 or 7$\times$4) without the substrate is $E^{scell-adsgeom}_{mol}$. The substrate energies, $E^{adsgeom}_{TMD}$ and $E^{relax}_{TMD}$, are the adsorbed and relaxed geometries, respectively (in their supercells, without pentacene). $E_{PEN/TMD}$ is the energy of the total heterostructure without missing components, and the system that defines adsorbed geometry.

\section{Density of States}

The highest occupied molecular orbital (HOMO) and lowest unoccupied molecular orbital (LUMO) of pentacene as determined by density of states (DOS) calculations are shown in Tables \ref{supp-tab-orbital-6x3-7x4-unrotated} and \ref{supp-tab-orbital-6x3-7x4}, for 0\textdegree\ and rotated geometry, respectively.

\begin{table}[H]
\centering
    \caption{HOMO and LUMO energy of pentacene, adsorbed on TMD substrate, without rotation. Energies in eV.}
    \begin{tabular}{ccccc}
    \hline
         TMD &  MoS$_2$ (0\textdegree) & MoSe$_2$ (0\textdegree) & WS$_2$ (0\textdegree) & WSe$_2$ (0\textdegree)\\
         \hline
         6$\times$3 & & & & \\
         HOMO & -4.821 & -4.823 & -4.829 & -4.782 \\
         LUMO & -3.721 & -3.673 & -3.679 & -3.682 \\
         \hline
         7$\times$4 & & & & \\
         HOMO & -4.788 & -4.738 & -4.747 & -4.744 \\
         LUMO & -3.638 & -3.638 & -3.597 & -3.594 \\
         \hline
         \hline
    \end{tabular}
    \label{supp-tab-orbital-6x3-7x4-unrotated}
\end{table}

\begin{table}[H]
\centering
    \caption{HOMO and LUMO energy of pentacene, adsorbed on TMD substrate, with rotation. Most favorable angle of rotation is given in parenthesis. Energies in eV.}
    \begin{tabular}{ccccc}
    \hline
         TMD & MoS$_2$ & MoSe$_2$ & WS$_2$ & WSe$_2$ \\
         \hline
         6$\times$3 & 0\textdegree & 0\textdegree & 60\textdegree & 60\textdegree \\
         HOMO & -4.821 & -4.823 & -4.830 & -4.787 \\
         LUMO & -3.721 & -3.673 & -3.680 & -3.637 \\
         \hline
         7$\times$4 & 60\textdegree & 60\textdegree & 60\textdegree & 60\textdegree \\
         HOMO & -4.788 & -4.739 & -4.747 & -4.744 \\
         LUMO & -3.638 & -3.639 & -3.647 & -3.594 \\
         \hline
         \hline
    \end{tabular}
    \label{supp-tab-orbital-6x3-7x4}
\end{table}

\begin{figure}[H]
    \centering
    \includegraphics[scale=0.2]{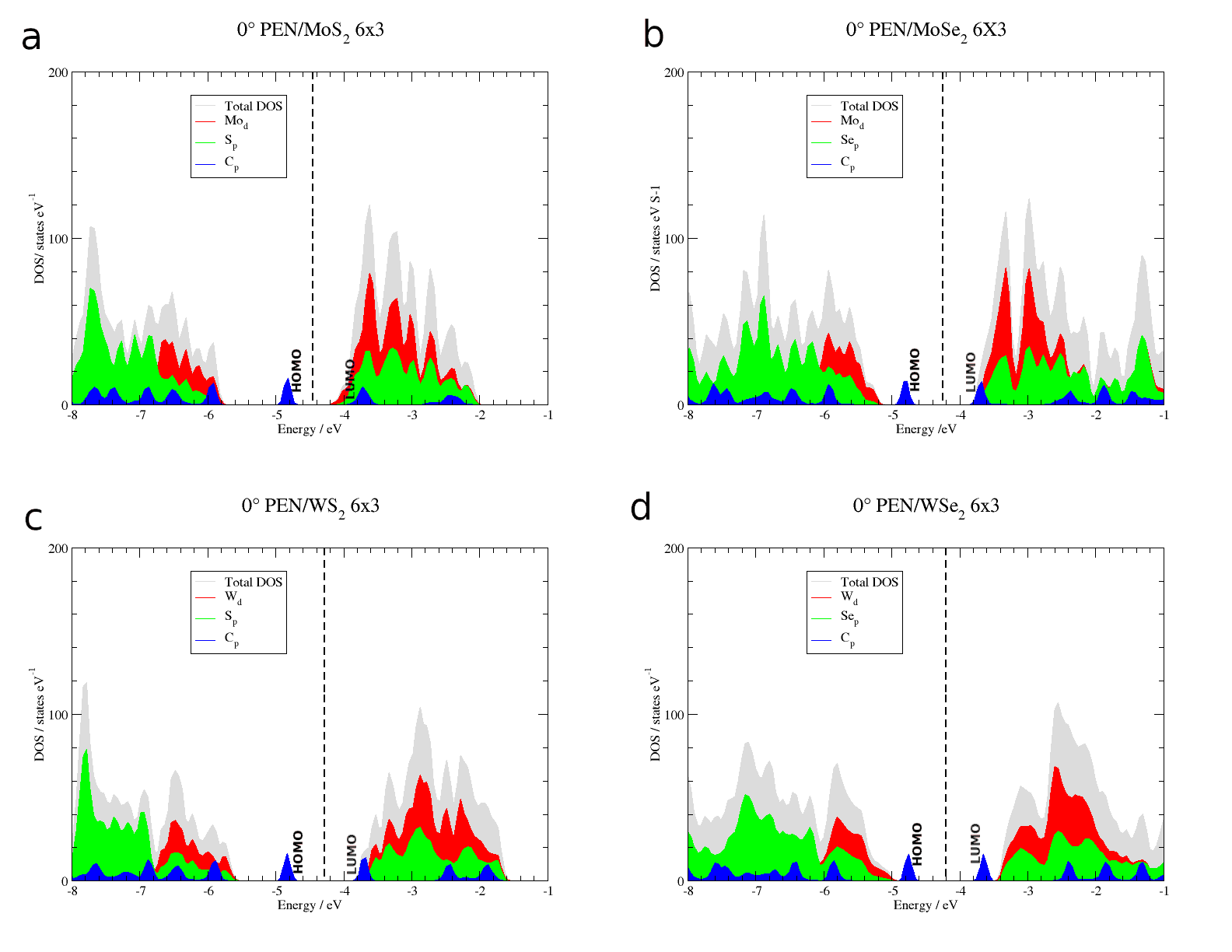}
    \caption{Total and projected density of states (DOS) of 0\textdegree\ (a) PEN/MoS$_2$, (b) PEN/MoSe$_2$, (c) PEN/WS$_2$ and (d) WSe$_2$ 6$\times$3 systems. Fermi energy is denoted by a vertical dashed line, and pentacene's HOMO and LUMO are labelled.}
    \label{supp-fig-unrotated-pDOS-6x3}
\end{figure}

\begin{figure}[H]
    \centering
    \includegraphics[scale=0.3]{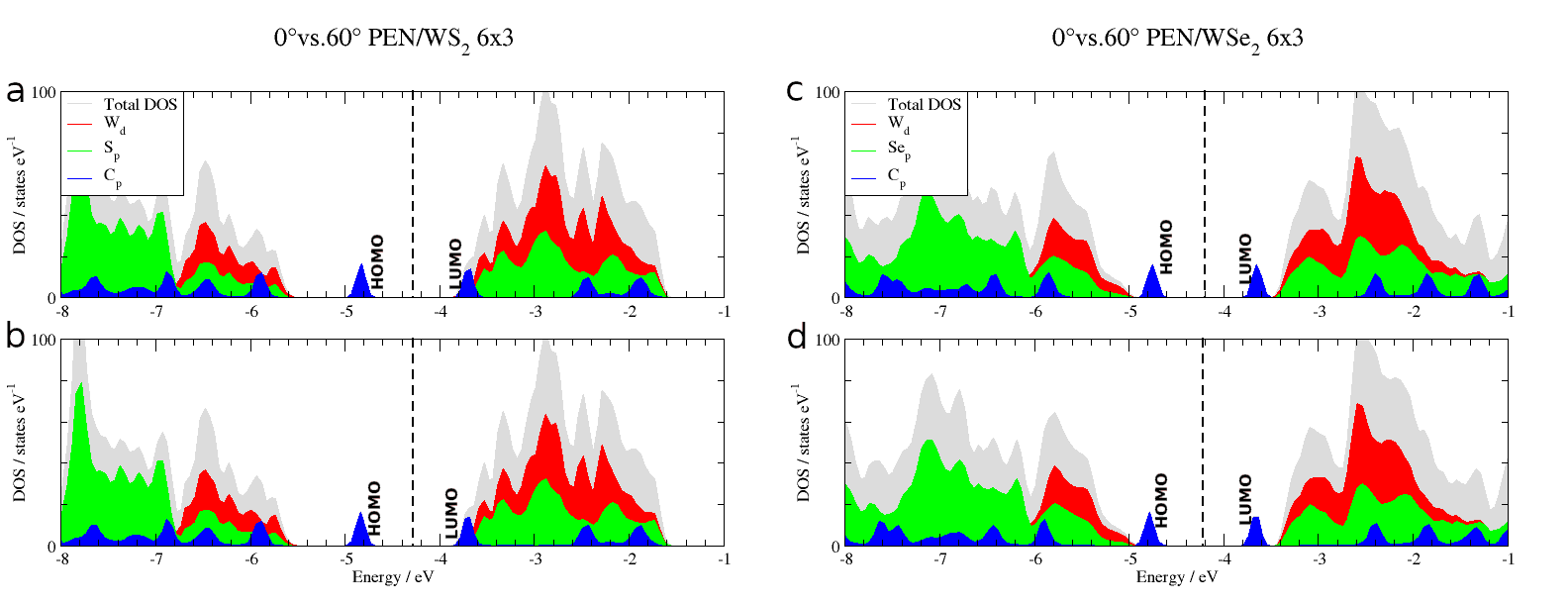}
    \caption{Total and projected density of states of 6$\times$3 systems, comparing WS$_2$ with 0\textdegree\ pentacene (a) and the energetically favorable pentacene rotation of 60\textdegree\ (b), and WSe$2$ with 0\textdegree\ pentacene (c) and the energetically favorable rotation of 60\textdegree\ (d).}
    \label{supp-fig-pDOS-comparison-6x3}
\end{figure}

The following figures show caparisons of the projected DOS of the 6$\times$3 systems, with Figure \ref{supp-fig-unrotated-pDOS-6x3} showing the 0\textdegree\ systems, and a comparison of the 0\textdegree\ with the favorably rotated tungsten systems (molybdenum systems favored 0\textdegree) in Figure \ref{supp-fig-pDOS-comparison-6x3}.

\begin{figure}[H]
    \centering
    \includegraphics[scale=0.2]{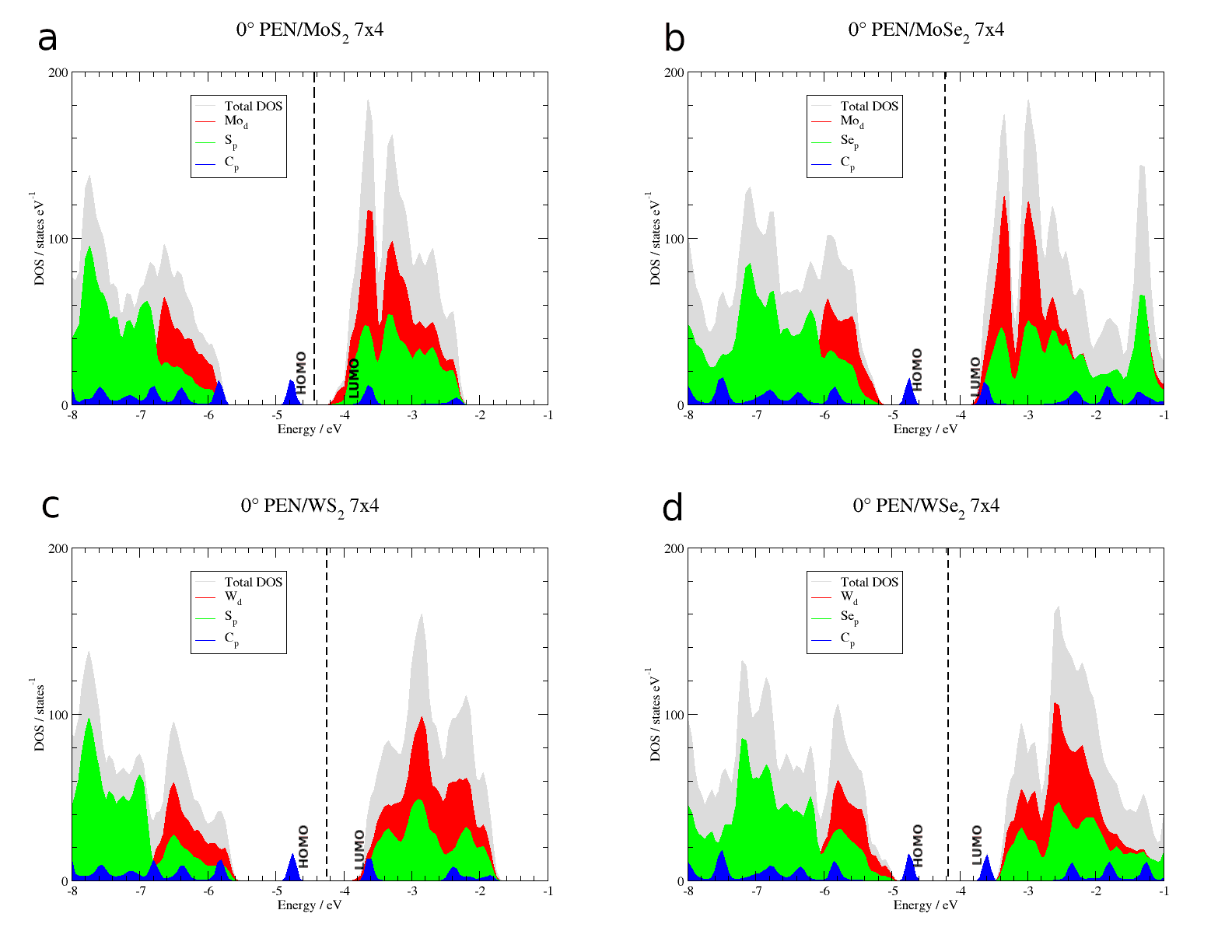}
    \caption{Total and projected density of states (DOS) of 0\textdegree\ (a) PEN/MoS$_2$, (b) PEN/MoSe$_2$, (c) PEN/WS$_2$ and (d) WSe$_2$ 7$\times$4 systems. Fermi energy is denoted by a vertical dashed line, and pentacene's HOMO and LUMO are labelled.}
    \label{supp-fig-unrotated-pDOS-7x4}
\end{figure}

\begin{figure}[H]
    \centering
    \includegraphics[scale=0.3]{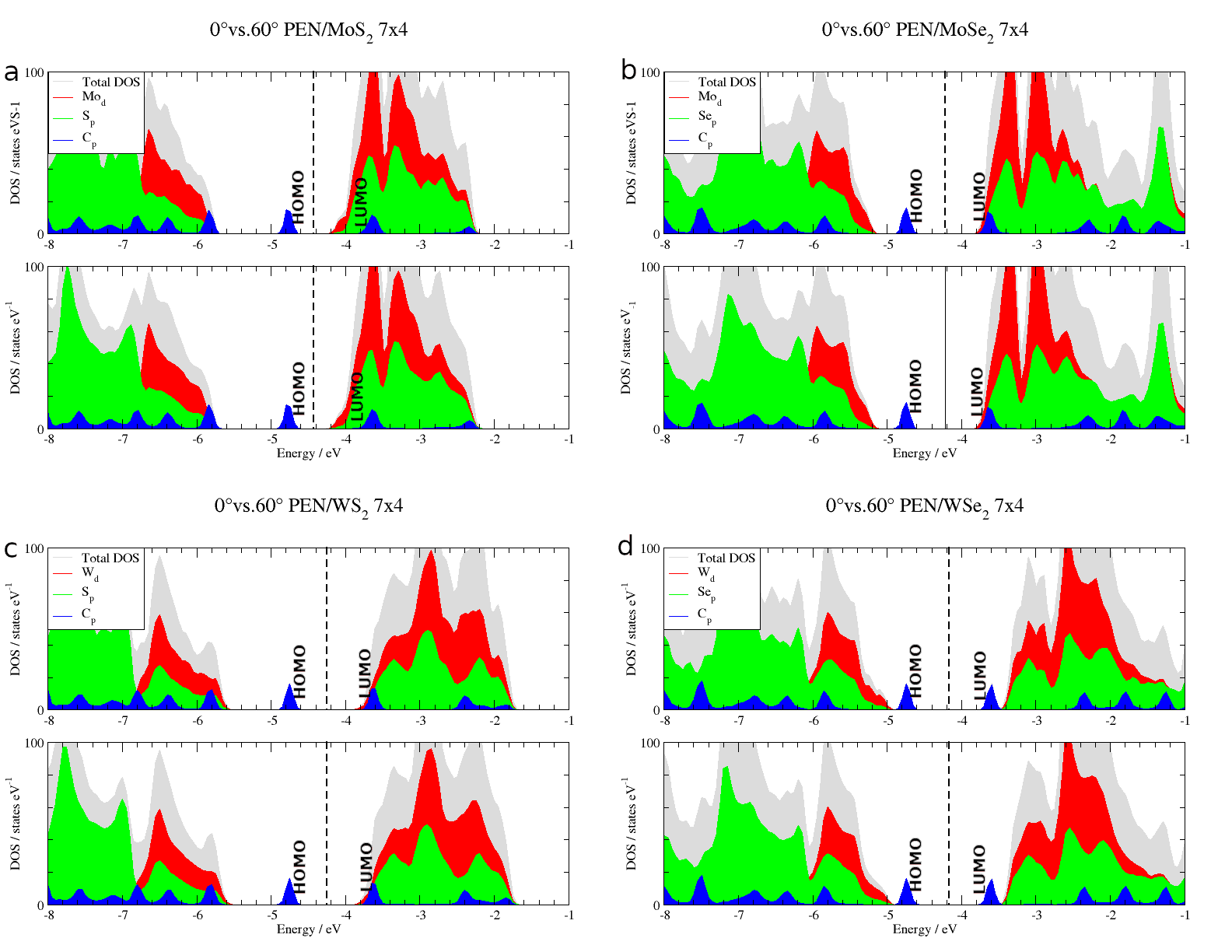}
    \caption{Total and projected density of states of 7$\times$4 PEN/MoS$_2$ (a) with 0\textdegree\ pentacene (top panel) and the energetically favorable pentacene rotation of 60\textdegree\ (bottom panel), PEN/MoSe$_2$ (b) with 0\textdegree\ (top panel) and 60\textdegree\ (bottom panel) pentacene, PEN/WS$_2$ (c) with 0\textdegree\ (top panel) and 60\textdegree\ pentacene (bottom panel), and PEN/WSe$_2$ (d) with 0\textdegree\ (top panel) and 60\textdegree\ (bottom panel) pentacene.}
    \label{supp-fig-pDOS-comparison-7x4}
\end{figure}

Figure \ref{supp-fig-unrotated-pDOS-7x4} shows the 0\textdegree\ 7$\times$4 systems, with a comparison of 0\textdegree\ systems with their rotated counterparts shown in Figure \ref{supp-fig-pDOS-comparison-7x4}.

\begin{figure}[H]
    \centering
    \includegraphics[scale=0.3]{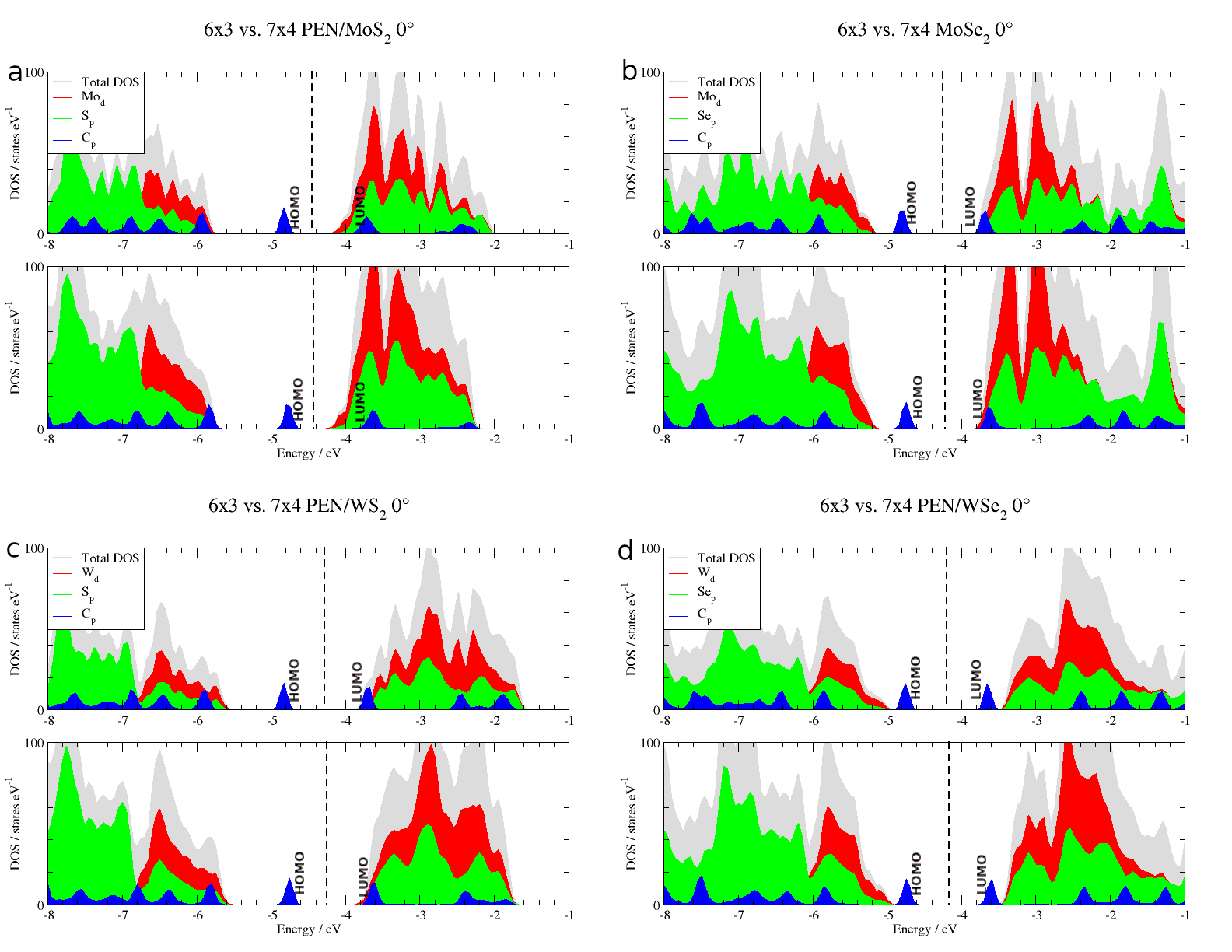}
    \caption{Total and projected density of states (DOS) of PEN/TMD systems with 0\textdegree\ pentacene, as a comparison of pentacene concentrations \cite{black2023interaction}: PEN/MoS$_2$ (a) in 6$\times$3 (top panel) and 7$\times$4 (bottom panel) supercells, PEN/MoSe$_2$ (b) in 6$\times$3 (top panel) and 7$\times$4 (bottom panel) supercells, PEN/WS$_2$ (c) in 6$\times$3 (top panel) and 7$\times$4 (bottom panel) supercells, and PEN/WSe$_2$ (d) in 6$\times$3 (top panel) and 7$\times$4 (bottom panel) supercells.}
    \label{supp-fig-pDOS-comparison-0}
\end{figure}

A comparison between concentration regimes with 0\textdegree\ pentacene is shown in Figure \ref{supp-fig-pDOS-comparison-0}.

\bibliography{references.bib}
\bibliographystyle{ieeetr}